# A Financial Risk Analysis: Does the 2008 Financial Crisis Give Impact on Weekend's Returns of the U.S. Movie Box Office?

Novriana Sumarti and Rafki Hidayat

*Abstract*— The Financial Crisis of 2008 is a worldwide financial crisis causing a worldwide economic decline that is the most severe since the 1930s. According to the International Monetary Fund (IMF), the global financial crisis gave impact on $3.4 trillion losses from financial institutions around the world between 2007 and 2010. Does the crisis give impact on the returns of the U.S. movie Box Office? It will be answered by doing an analysis on the financial risk model based on Extreme Value Theory (EVT) and calculations of Value at Risk (VaR) and Expected Shortfall (ES). The values of VaR and ES from 2 periods, 1982 – 1995 and 1996 – 2010, are compared. Results show that the possibility of loss for an investment in the movie industry is relatively lower than the possibility of gain for both periods of time. The values of VaR and ES for the second period are higher than the first period. We are able to conclude that the 2008 financial crisis gave no significant effect on these measurement values in the second period. This result describes the high potential opportunity in the investment of the U.S. movie makers.

## I. INTRODUCTION

The Financial Crisis of 2008, continuing into 2009, is a worldwide financial crisis triggered by bad investment decisions by major banks in the U.S. and Europe on potentially unpayable mortgages. Globally the worst condition occured around September 2009, and the major impact was felt by people in Britain, European Union, Russia, Japan, the oil countries of the Middle East, and the Third World. The crisis affected virtually every sector of the economy, including housing, construction, office buildings, automobiles, retail sales, and government [2]. According to the International Monetary Fund (IMF) [1], the global financial crisis gave impact on $3.4 trillion losses from financial institutions around the world between 2007 and 2010. However, were the returns of the U.S. movie Box Office in that period also affected by the crisis?

In [6] and [7], De Vany and Walls discovered that the movie industries are very risky. Half the theater screens in the United States are in bankruptcy and only 18% of movies earn a profit. Walls [8] analyzed on mobster-related movies which their profitability prospects are computed from the fitted Lévy-stable distribution. Using the stable Paretian probability model to capture the investment's uncertainty, De Vany [9] investigated that the probabilities of motion picture outcomes are much different from the Normal distribution. The tails of the distribution are "heavy" and large-scale events are much probable in a Paretian than in a Gaussian model. His paper claims that the variance is infinite and, for some variables, even the mean does not exist. Movie box office revenues, therefore, have no natural size or scale and there is no typical or average movie.

However in spite of the above gloomy picture, some movie makers took risk to make high budget projects and expected in return the movie will be categorized as blockbuster films. The U.S. weekend box office revenue is dominated by high budget movies. According to the Internet Movie Database (IMDB), among 360 blockbusters with gross box office income of over $100 million during their theatrical runs, 290 movies, or about 80%, had budgets above $60 millions . At most cases, the distribution of box office revenue is dominated by these high budget movies. However, this is not always the case. Some high budget movies may sustain losses at the box office [14]. An analysis of the financial risk on this revenue is needed in order to measure the potential opportunity in the investment of the movies categorized as the box office movies.

The main purpose of a financial risk model is to make forecasts of the likely losses or gains that would be incurred for a variety of risks so it can be as references of the actual or potential investment. In engineering practise, a risk assessment is well-established but derived from different approach. A risk assessment is a careful examination of potential causes to harm to people, so it can analyses whether enough precautions has ben taken to prevent harm [10]. Using algebraic modelling, the mathematical derivation of the risk assessment applied on technological systems had been conducted in [11].

In this paper we make use of Extreme Value Theory (EVT) [3,4,5], and Value at Risk (VaR) which was introduced by JP Morgan in the mid 1990s who launched the RiskMetrics methodology [12] and formulized theoretically by Artzner, Delbean, Eber and Heath [3,4]. The latter papers concerned about VaR being used as a risk measure but it is not necessarily subadditive or *incoherent*, which means there is no guarantee that merger of two portfolios do not create extra risk. In fact, diversification of portfolio, containing more than one asset, can reduce risk. Another alternative measures introduced later is Expected Shortfall (ES) which is a coherent measure [4]. Using these risk measures, some research were conducted on financial market [12], actuarial/insurance data [13,14] and movie industry [15,16]. The latter papers use the same kind of data and similar tools but different approach with this paper, so a significantly different result will be given in this paper.

This paper analyzes the return of weekend's earnings of the US movie Box Office taken from January 1982 to December 2010. Based on EVT analysis, only extreme (positive and negative) values of data are examined, those are data surpassing a certain threshold $u$. These negative and positive tails of the data will be fitted into a Generalized Pareto distribution model with particular values of parameters $\xi$ and $\sigma$. In the next step, values of VaR and ES are computed from 2 periods, 1982 – 1995 and 1996 – 2010. Note that although VaR is an incoherent measure, it satisfies monotonicity condition [4]: for X and Y are different portfolios, if $X \leq Y$, $VaR(X) \leq VaR(Y)$. The analysis is comparing the results and trying to answer a question whether the 2008 financial crisis has given an effect on the returns of the U.S movie Box Office or not. The less-extensive version of this work has been published in [15].

In the next section, we will explain in detail the definitions and techniques we used in EVT. In section 3 we will discuss how to choose an optimal threshold $u$ in the Generalized Pareto distribution mode. In section 4 the data of the weekend's revenue of the US movie Box Office will be examined using Value at Risk (VaR) and Expected Shortfall (ES). The conclusion will be written in section 5.

## II. EXTREME VALUE THEORY

The extreme value theory (EVT) is the name of methods for modeling and measuring extreme risks. For example, it concerns the determination of the risk for the losses or gains of the weekend's returns of the US movie Box Office due to uncertainty aspects of the movie industry. The potential values of a risk have a probability distribution which we will never know exactly. We could use past losses due to similar risks, which may provide partial information about that distribution. We develop a model by selecting a particular probability distribution. We may have estimated this distribution through statistical analysis of empirical data. We could develope a model based on the Normal distribution, which is commonly used for large amount of data. However, extreme events occur when a risk takes values from the tail of its distribution. On the other hand, the tails of the Normal distribution are too thin to address the extreme loss, so the model does not provide much information about the extreme risk. EVT is a tool which attempts to provide us with the best possible model of the tail area of the distribution, rather than of all existing data.

In Figure 1, from [5,12] we summarize a classification of models included in EVT. The block maxima is a model for the largest observations collected in a particular "continuous" period, for instance daily or hourly, from large samples of identically distributed observations. These observations will be classified into some blocks and the maximas from these blocks are considered as variables randoms which will be fitted into a distribution function. The other model, the peaks-over-threshold (POT) model, is a model for all large observations which exceed a high threshold. In this paper we use the POT models, which are considerably as the most useful for practical applications due to the data are not necessary identical distributed.

Within the POT class of models, there are the semi-parametric models built around the Hill estimator and its relatives, and the fully parametric models based on the generalized Pareto distribution or GPD. When used correctly, both are theoretically justifed and empirically useful. The latter method is used in this paper because it obtains simple parametric formulae for measures of extreme risk.

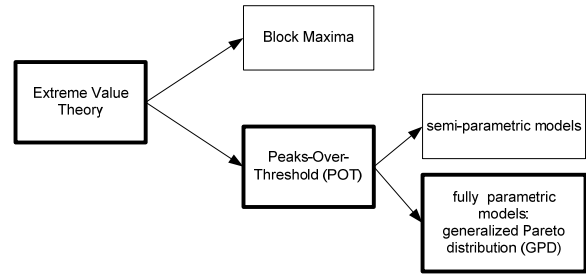

Figure 1    Classification of EVT methods

Having estimated a distribution model, we calculate the quantile risk measure or Value at Risk (VaR). For a given probability and a time horizon, VaR assesses the worst-case loss, where the worst case is defined as the event with a *1-p* probability. It can also be applied to the `best case' gain. VaR is defined as a threshold value such that the probability is in the given probability level. The value VaR is $VaR_p = F^{-1}(1-p)$, $F$ is the cdf of the loss distribution and $p$ is the given probability level. $VaR_p$ is the $p$th quantile of the distribution F. In this paper it will be estimated by function

$$VaR_p = u + \frac{\sigma}{\xi}\left(\left(\frac{n}{n_u}p\right)^{-\xi} - 1\right), \qquad (1)$$

where $u$ is the chosen threshold, $n$ is the total number of data, and $n_u$ is the number of data surpassing the threshold $u$. Parameter $u$ is obtained from the selection of threshold in POT, parameters $\xi$ and $\sigma$ are obtained from the application of Maximum Likelihood Estimation (MLE). Those parameters which will be explained in the next section.

The quantile risk measure does not take into consideration what the loss/gain will be if that 1-*p* worst/best case event actually occurs. The expected shortfall or conditional tail expectation is defined as the expected size of a return that exceeds $VaR_p$ or $ES_p = E(X|X > VaR_p)$. The formula for estimated it is

$$ES_p = \frac{VaR_p + \sigma - u\xi}{1-\xi}. \qquad (2)$$

Having the values $VaR_p$ and $ES_p$ from the weekend's returns of the US movie Box Office, we can conclude the potential investment of US movie industry. The works in [15] and [16] had discussed this topic but we use quite different tools and analysis. For example, instead of logaritmic difference, we use return formulae defined in (3). This usage of return gives a significant different between positive and negative returns, especially in their Mean-Excess plots, which is not shown in the existing papers.

In the next section we discuss tools to choose an optimal threshold $u$ using statistics quantities $W^2$ and $A^2$, and the graphical observation of the mean-excess function plots. With this threshold, the estimation values of of parameters $\xi$ and $\sigma$ can be found using Maximum Likelihood Estimation (MLE). It is the method of selecting values of the model parameters that maximize the GP pdf function as the likelihood function. We also use another approach to choose an optimal threshold $u$ for both positive and negative returns explained in the subsequently section.

## III. OPTIMAL THRESHOLD DETERMINATION

Data from the weekend's earning of the US movie Box Office is presented in the form of their return, the relative difference between values of the return from a week and of the previous week, and denoted as a random variable $X$.

$$X_i = \frac{R_{i+1} - R_i}{R_i}, i = 1, \ldots, n. \quad (3)$$

Now using the Peaks over Threshold (PoT) method, the observed data are surpassing a certain threshold u and will be the negative and positive tails of the whole data. For a certain value of $u$, we need to model the distribution of conditional excess $F_u$ which is

$$F_u(y) = P(X - u \le y \mid X > u), 0 \le y \le x_F - u, \quad (4)$$

where $y = x - u$ are the excesses, and $x_F$ is the right endpoint of data. This distribution will be fitted to a Generalized Pareto Distribution (GPD) model for specific values of its parameters because, according to the Pickands, Balkema and de Haan (PBH) theorem ([18],[19]), function $F_u(y)$ for large value of $u$ is well approximated by the GPD, that is

$$G_{\xi,\sigma}(y) = \begin{cases} 1 - \left(1 + \frac{\xi}{\sigma} y\right)^{-1/\xi} & \text{if } \xi \ne 0 \\ 1 - e^{-y/\sigma} & \text{if } \xi = 0 \end{cases} \quad (5)$$

for $y \in [0, x_F - u]$ if $\xi \ge 0$ and $y \in \left[0, -\frac{\xi}{\sigma}\right]$ if $\xi < 0$, where $\xi$ is a shape parameters and $\sigma$ is a scale parameter. Note that this distribution was developed as a distribution that can model tails of a wide variety of distributions. In the case $\xi > 0$ GPD is heavy-tailed distribution, and if $\xi < 0$ GDP is short-tailed distribution. Maximum likelihood estimation (MLE) can be used to estimate the parameters of the GPD, $\xi$ and $\sigma$, which are the solution of the system

$$\frac{1}{\xi^2} \left( \sum_{i=1}^n \ln(\sigma + \xi y_i) - n \ln(\sigma) \right) - \left(\frac{1}{\xi} + 1\right) \left( \sum_{i=1}^n \frac{y_i}{\sigma + \xi y_i} \right) = 0,$$

$$\frac{1}{\sigma}(1 + \xi) \left( \sum_{i=1}^n \frac{y_i}{\sigma + \xi y_i} \right) - \frac{n}{\sigma} = 0,$$

for $\xi \ne 0$, and $\sigma = \frac{1}{n} \sum_{i=1}^n y_i$ for $\xi = 0$. The system of equations above comes from the first derivatives of the log-likelihood function for the GPD for a sample $y = \{y_1, y_2, \ldots, y_n\}$ of exceess.

How to choose an optimal threshold $u$ is a key problem that decides the fraction of data belonging to the tail, and therefore affects the results of the MLE of the parameters of the GPD function. There is a trade-off between the condition that the value of $u$ should be high enough to satisfy the PBH theorem, and the fact that the higher the threshold, the fewer observations are left for the estimation of the parameters. Two simple tools among other approaches are the goodness-of-fit test procedure using $W^2$ and $A^2$, and the graphical plot of the empirical mean excess function.

From [13], the goodness-of-fit test using Cramer-von Mises statistics $W^2$ and the Anderson-Darling stastic $A^2$ are adopted as a tool to proof a nul hypothesis saying that the random sample $X_1, X_2, \ldots, X_n$ come from GPD model (4). We transform them into $Z_i = G_{\xi,\sigma}(X_i)$ by sorting the order of $X_1, X_2, \ldots, X_n$, and then estimate values of $\xi$ and $\sigma$ using MLE. Quantities below are calculated using only nonzero values of $Z_i$:

$$W^2 = \sum_{i=1}^n \left(z_i - \frac{2i-1}{2n}\right)^2 + \frac{1}{12n},$$
$$A^2 = -n - \frac{1}{n} \sum_{i=1}^n (2i-1)[\log(z_i) + \log(1 - z_{n+1-i})].$$

Those values of $W^2$ and $A^2$ are compared to a table of critical values under a given significant level of $\alpha$, which can be seen in Table 1 below. The completed table can be found in [13]. Note that the table is provided only for positive values of $\xi$.

Table 1 Goodness-of-fit test values for GPD model
$\alpha = \Pr(W^2 \ge z)$ or $\alpha = \Pr(A^2 \ge z)$

| $\xi \backslash \alpha$ | | 0.500 | 0.25 | 0.100 | 0.050 | 0.025 | 0.010 | 0.005 |
|---|---|---|---|---|---|---|---|---|
| 0.00 | $W^2$ | 0.057 | 0.086 | 0.124 | 0.153 | 0.183 | 0.224 | 0.255 |
|      | $A^2$ | 0.397 | 0.569 | 0.796 | 0.974 | 1.158 | 1.409 | 1.603 |
| 0.10 | $W^2$ | 0.055 | 0.081 | 0.116 | 0.144 | 0.172 | 0.210 | 0.240 |
|      | $A^2$ | 0.386 | 0.550 | 0.766 | 0.935 | 1.110 | 1.348 | 1.532 |
| 0.20 | $W^2$ | 0.053 | 0.078 | 0.111 | 0.137 | 0.164 | 0.200 | 0.228 |
|      | $A^2$ | 0.376 | 0.534 | 0.741 | 0.903 | 1.069 | 1.296 | 1.471 |
| 0.30 | $W^2$ | 0.052 | 0.076 | 0.108 | 0.133 | 0.158 | 0.193 | 0.220 |
|      | $A^2$ | 0.369 | 0.522 | 0.722 | 0.879 | 1.039 | 1.257 | 1.426 |

In the second tool, the plot of the Empirical Mean Excess function $e(u) = E(X - u \mid X > u)$ is calculated as in [15,16]:

$$e(u) = \frac{\sigma + \xi u}{1 - \xi}, \text{ for } \xi < 1, \sigma + \xi u > 0. \quad (6)$$

The mean excess function $e(u)$ can be estimated using the following sample empirical mean excess function

$$e_n(u) = \frac{\sum_{i=1}^n (x_i - u)}{\sum_i^n 1_{\{x_i > u\}}} \quad (7)$$

where $\sum_i^n 1_{\{x_i > u\}}$ is the number of data surpassing the value of $u$. From equation (6), $e(u)$ is a positive linear function of $u$ so is $e_n(u)$. A downward trend where $\xi < 0$ is an indication of a short-tailed distributions. On the contrary, an upward trend where $\xi > 0$ is for heavy-tailed behavior [13]. According to [16], we have to select $u$ that located at the beginning of a portion of the sample mean excess plot that is roughly linear and sloping up. Unfortunately, there are many potential values of $u$ resulted from this technique and the goodness-of-fit test. In the next section, we use additional approach by finding the maximum values of VaR and ES, and then doing the analysis to answer our question in the beginning of this paper.

## IV. DATA ANALYSIS

Figure 2 shows the yearly average returns of the weekend's earnings taken from www.boxofficemojo.com. There will be two periods of time where the risk model will be calculated. The first period is 1982-1995 and the second is 1996-2010. In this figure, the graph from the second period shows an extreme fluctuation which will be examined whether this phenomenon can be captured by the calculation of values of VaR or not. Especially year 2008 is commonly known as the beginning of the financial crisis which gives bad impacts to virtually all sectors. Here the linear trendline shows straight line with nearly zero slope.

Data contains positive and negative returns which determine gains and losses. Graph in Figure 2 is above the horisontal axis which means the gains outnumbers the losses. It is also shown by box-plots for those two periods in Figure 3 using [20]. If IQR (interquartile range) is the difference between third quartile (Q3) and first quartile (Q1), the ends of the whisker are set at 1.5*IQR above Q3 and 1.5*IQR below Q1. Outliers are the minimum or maximum values outside this range, but only the min and max outliers are shown in the box-plot.

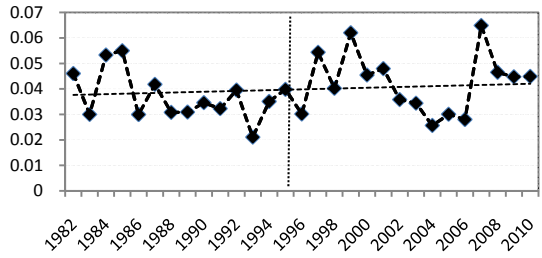

Figure 2    Average returns of the weekends's return.

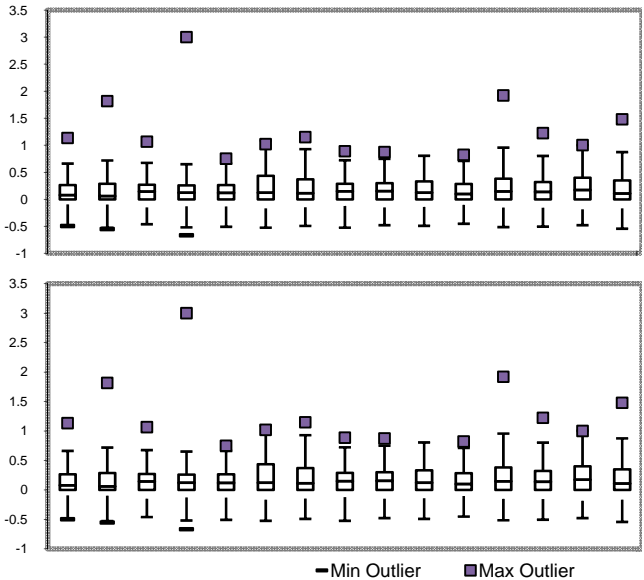

Figure 3  Box plots of returns from 1982-1995 (up) and and 1996-2010 (down).

There is always max (positive) outliers every yearly returns, but only few have min (negative) outliers. It can be predicted there will be heavy tails for positive returns and short tails for negative returns.

Now for each period, the data is divided into positive and negative returns as their distributions are asymmetric. The observed sample comprises 1575 data, consisting 358 positive returns and 371 negative returns for period 1982-1995, and also 407 positive returns and 439 negative returns for period 1996-2010.

*A. Positive Returns*

For the first period, the mean-excess of the positive returns using formula (7) is plotted in Fig.4. The values $u = 0.04125$ and $u = 0.08479$ in the graph are chosen using the technique explained shortly. The graph tends to sloping up which means that formula (6) has positive value of $\xi$. It means a heavy-tailed behavior occurs for the positive returns in this period, which confirms the prediction from the blox-plots. From this fact, we will only consider values of $u$ that have corresponding positive values of $\xi$ coming from the MLE computation. There are 339 values of positive $\xi$ and denoted by index 1 to 339 based on values of $\xi$ from the smallest to the largest.

To select $\xi$ in order to have an appropiate measure of risk representing the period of time being observed, we choose the value which give the maximum value of VaR calculated from equation (1). The latter will be compared to the max value from the other period. For *p = 0.01* Figure 5 shows the VaR values for all indexes. We computed them only for *p = 0.01* because it distinguised significantly between those two periods.

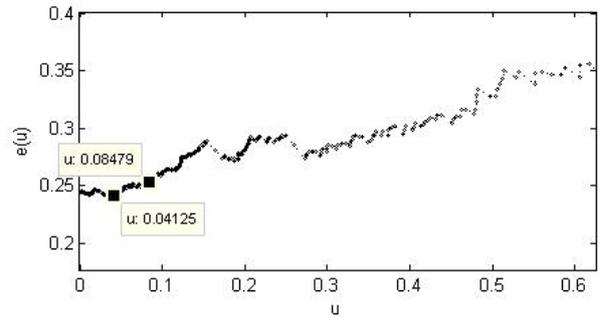

Figure 4   Mean-excess plot for positive returns 1982-1995.

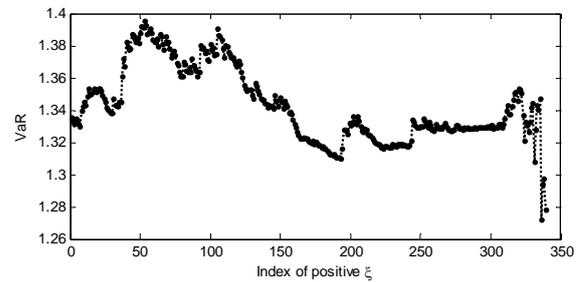

Figure 5  The values of VaR with p=1%  for positive returns 1982-1995.

The max VaR value is 1.3954, and its value of ES calculated from equation (2) is 1.9376. The corresponding value of $u$ is 0.04125 with $\xi = 0.1814$ and $\sigma = 0.1982$. If we calculate the goodness-of-fit test for GPD model, it has $W^2 = 0.13099$ and $A^2 = 0.74268$. Based on Table 1, the null hypotesis is accepted with $\alpha = 0.25$. Notice that the value of $\alpha$ is not determined beforehand but it comes from the chosen maximum value of VaR.

If we decide to have the smaller value of $\alpha$ in order to accept the null hypotesis, choose commonly $\alpha = 0.05$, there are 23 values of $u$ and the max VaR occurs at $u = 0.08479$, $\xi = 0.2037$ and $\sigma = 1.0455$.

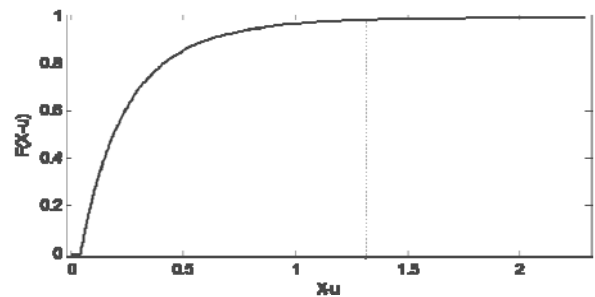

Figure 6   The *1%*-th quantile of the distribution *F* on 1982-1995 positive returns $u = 0.04125$.

What do the values of VaR and ES tell us? For positive returns $u = 0.04125$, it show that, with 1% probability, the returns from one weekend to the next could exceed 139.54%, and the average returns above this level will be 193.76 %. This is a very good result. In fact, there are 9 of 358 returns data in the first period giving values of more than 100%, and the maximum value is 233.52 %. Figure 6 gives the description of the *1%*-th quantile of the distribution *F*.

We do the same technique to give results for the second period, 1996-2010. In Figure 7, the mean-excess plot of the positive returns presents the same behaviour with the related plot for the previous period, where the graph tends to sloping up, so a heavy-tailed behavior occurs. The chosen value $u = 0.24240$ with $\alpha = 0.10$ gives value 1.4718 as the max VaR of this period. It means that with 1% probability, the returns from one weekend to the next could exceed 147.18%, and the average returns above this level will be 218.19 %. These values are higher than of the previous period. Data shows 12 of 407 values of returns exeeding 100% and the max value is 300%. So it is no surprise that the analysis yields the higher values of Var and ES. Using the other values of u for $\alpha = 0.05$, the values of VaR and ES are also higher than of the preceeding period of time. These results are shown in Table 2.

on their max values which are written in Table 3.

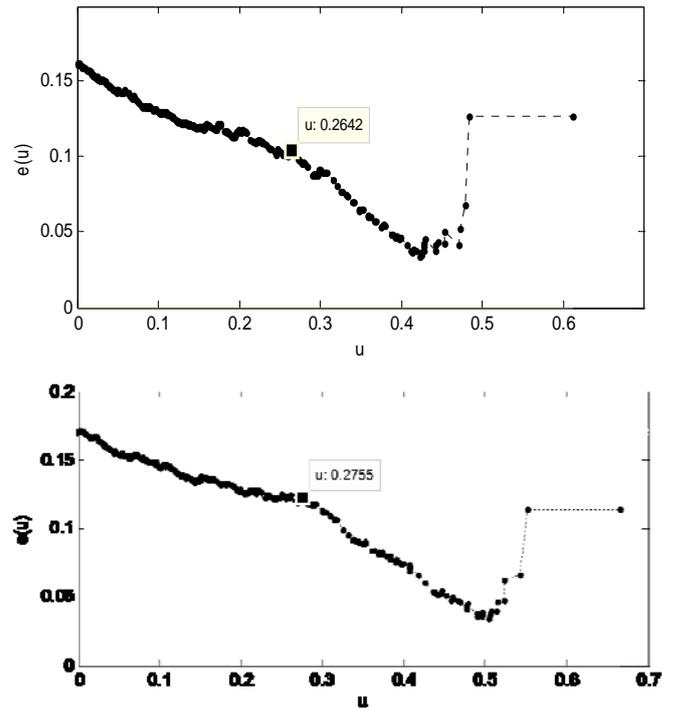

Figure 8 Mean-excess plots for negative returns in 1982-1995 (up) and in 1996 – 2010 (down).

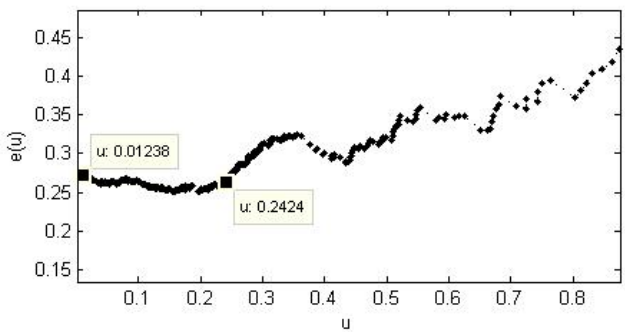

Figure 7 Mean-excess plot for positive returns 1996-2010.

Table 2 Numerical results for positive returns

|  | 1982-1995 | | 1996-2010 | |
| --- | --- | --- | --- | --- |
|  | $\alpha = 0.25$ | $\alpha = 0.05$ | $\alpha = 0.10$ | $\alpha = 0.05$ |
| u | 0.04125 | 0.08479 | 0.24240 | 0.01238 |
| ξ | 0.1814 | 0.2037 | 0.2658 | 0.2061 |
| σ | 0.1982 | 1.0455 | 0.1946 | 1.1439 |
| VaR | 1.3954 | 0.3747 | 1.4718 | 0.8333 |
| ES | 1.9376 | 1.8802 | 2.1819 | 1.1641 |

*B. Negative Returns*

Now we discuss the negative returns of both periods. The mean-excess plots using formula (7) for both periods have the same behaviour as it is shown in Figure 8. Here the graphs tend to sloping down which means that GDP formula (6) has negative value of ξ, so values of *u* being considered have negative values of ξ resulted from the MLE computation. The negative value is an indication of a short-tailed distributions, which also confirms the prediction from the blox-plots. We cannot use the Goodness-of-fit test using table 1 because it does not cover negative values of **ξ.** Now to choose the optimal values of *u*, we calculated VaR values for p = 0.01. In Fig. 9, the plots of VaR values for both periods of time are presented. All values of VaR for the first period is smaller than of the second period. This also reflects

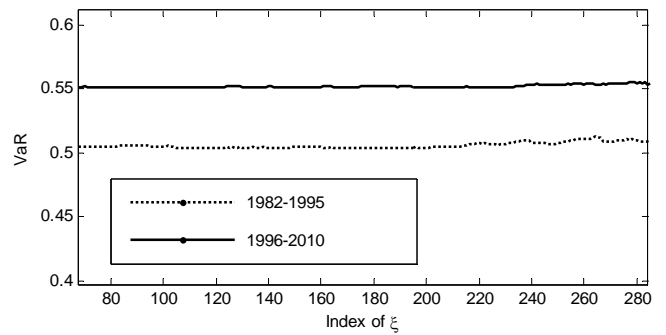

Figure 9 The values of VaR with p=1% for negative returns.

Table 3 Numerical results for negative returns

|  | 1982-1995 | 1996-2010 |
| --- | --- | --- |
| u | 0.26423 | 0.2755 |
| ξ | -0.3586 | -0.3974 |
| σ | 0.1374 | 0.1664 |
| VaR | 0.5130 | 0.5647 |
| ES | 0.5485 | 0.6016 |

Those values means there is 1% probability that the loss from one weekend to the next could exceed 51.30% in period 1982-1995 and 56.47% in period 1996-2010. It shows higher loss in the latter period although the values are not significantly different. The average loss above the corresponding VaR levels will be 54.85% (1982-1995) and 60.16% (1996-2010).

## V. CONCLUSIONS

Model of the tails of the GDP distribution for returns of weekend movie box office in the US using extreme value theory shows heavy-tails for the positive returns, and short-tails for the negative returns. The positive and negative returns from period 1982-1995 have higher VaR values than of period 1996-2010, which means the possibility of gain for an investment in the movie industry in the latter period is higher than the possibility in the earlier period. However, the possibility of loss in the latter period is also higher than of in the earlier period. It can be concluded that 1998 financial crisis does not give impact on the investment in the blockbuster movie industry. In further work, we could measure the risk of the investment in the overall returns of movie industry, not only for the blockbuster movies, which is already proven in this paper a good investment in entertainment industry.

## ACKNOWLEDGMENT

We thank Dr. K.I.A. Syuhada for helping the initial work, Prof. D.E. Giles for email correspondences at the beginning of the research, and participants at ICAEM-WCE 2011 for good inputs during the presentation.